\documentclass[12pt]{article}
\usepackage{bbm}
\usepackage{epsfig}
\newcommand{\be}{\begin{equation}}
\newcommand{\ee}{\end{equation}}
\newcommand{\ba}{\begin{eqnarray}}
\newcommand{\ea}{\end{eqnarray}}

\newcommand{\lesssim}{\:\mbox{\raisebox{-3pt}{$\stackrel%
{\displaystyle <}{\sim}$}}\:}
\newcommand{\gtrsim}{\:\mbox{\raisebox{-3pt}{$\stackrel%
{\displaystyle >}{\sim}$}}\:}
\begin{document}
\title{\normalsize \hfill UWThPh-2003-42 \\[1cm] \LARGE
Leptogenesis in seesaw models
with a \\
twofold-degenerate neutrino Dirac mass matrix}
\author{Walter Grimus\thanks{E-mail: walter.grimus@univie.ac.at} \\
\setcounter{footnote}{6}
\small Institut f\"ur Theoretische Physik, Universit\"at Wien \\
\small Boltzmanngasse 5, A--1090 Wien, Austria \\*[3.6mm]
Lu\'\i s Lavoura\thanks{E-mail: balio@cfif.ist.utl.pt} \\
\small Universidade T\'ecnica de Lisboa \\
\small Centro de F\'\i sica das Interac\c c\~oes Fundamentais \\
\small Instituto Superior T\'ecnico, P--1049-001 Lisboa, Portugal \\*[4.6mm] }

\date{10 May 2004}

\maketitle

\begin{abstract}
We study leptogenesis in two seesaw models
where maximal atmospheric neutrino mixing and $U_{e3} = 0$ result from
symmetries. Salient features of those models
are the existence of three Higgs doublets and 
a twofold degeneracy of the neutrino Dirac mass matrix. 
We find that in those models 
both leptogenesis and neutrinoless double beta decay
depend on the same unique Majorana phase.
Leptogenesis can produce a baryon asymmetry of the universe of the right size
provided the mass of the heavy neutrino whose decays generate the
lepton asymmetry is in the range
$10^{11}$--$10^{12}\, \mathrm{GeV}$.
Moreover, in these models, leptogenesis precludes an inverted neutrino
mass spectrum since it requires the mass of the lightest neutrino to
be in the range $10^{-3}$--$10^{-2} \, \mathrm{eV}$. 
\end{abstract}

\newpage

\section{Introduction}
\label{introduction}

Experimental cosmology has witnessed spectacular progress
during the last few years.
In particular,
the WMAP experiment \cite{WMAP} has determined with fantastic precision
the baryon asymmetry of the universe,
which is given by the ratio of baryon number over the number of photons,
experimentally measured to be
\be\label{etaBexp}
\eta_B \equiv \frac{n_B - n_{\bar B}}{n_\gamma}
= 6.5^{+0.4}_{-0.3} \times 10^{-10},
\ee
where $n_B$ is the (present) baryon density
of the universe,
$n_{\bar B}$ is the anti-baryon density
and $n_\gamma$ is the (present) density of photons.
This value of $\eta_B$
is in excellent agreement with the one inferred
from big bang nucleosynthesis \cite{steigman}.

Another field which has lately witnessed outstanding experimental progress
is neutrino masses and lepton mixing.
On the one hand,
the first results of the KamLAND experiment \cite{KamLAND}
have conclusively proved that solar neutrinos oscillate;
a global analysis of all solar neutrino results,
including the recent SNO measurement \cite{SNO} and also the KamLAND result,
gave 
a mass-squared difference \cite{SNO} 
\be\label{dm2solarexp}
\Delta m^2_\odot \equiv m_2^2 - m_1^2
= 7.1^{+1.2}_{-0.6} \times 10^{-5}\, {\rm eV}^2
\ee
and a large but non-maximal mixing angle 
\be\label{thetaexp}
\theta = 32.5^{+2.4}_{-2.3} \; \mbox{degrees},
\ee
where the errors reflect $1 \sigma$ constraints
in the two-dimensional $\theta$--$\Delta m^2_\odot$ region.
Note that $\tan{\theta}$ is the ratio,
in the decomposition of the electron neutrino $\nu_e$,
of the amplitude for the heavier neutrino $\nu_2$
over the amplitude for the lighter neutrino $\nu_1$:
$\tan{\theta} = \left| U_{e2} \left/ U_{e1} \right. \right|$,
where $U$ is the lepton mixing matrix.
On the other hand,
the Super-Kamiokande experiment \cite{SK}
has shown that atmospheric neutrinos
oscillate with a mass-squared difference in the 90\% CL range 
\be\label{dm2atm}
1.3 \times 10^{-3}\, \mathrm{eV}^2 < 
\Delta m^2_{\rm atm} \equiv \left| m_3^2 - m_1^2 \right|
< 3.0 \times 10^{-3}\, \mathrm{eV}^2,
\ee
with best-fit value 
$\Delta m^2_{\rm atm} = 2.0 \times 10^{-3}\, \mathrm{eV}^2$, 
and a (most likely) \emph{maximal} mixing angle:
\be
\sin^2{2 \theta_{\rm atm}} \equiv 4 \left| U_{\mu 3} \right|^2
\left( 1 - \left| U_{\mu 3} \right|^2 \right) > 0.90
\label{utbaz}
\ee
at 90\% CL,
the best-fit value being exactly 1.
Finally,
the CHOOZ experiment \cite{CHOOZ} and all other neutrino oscillation
data yield the upper bound 
$\left| U_{e3} \right|^2 \lesssim 0.054$ at $3 \sigma$ \cite{maltoni}.
For recent reviews on neutrino oscillations see \cite{reviews}.

These experimental developments invite a renewed interest of theorists
for leptogenesis \cite{yanagida,lepto-reviews}.
This is the possibility that the baryon asymmetry of the universe
has been generated through the standard-model sphaleron transmutation
of a previously existing lepton asymmetry,
which in turn was generated at the decay
of the heavy neutrinos involved in the seesaw mechanism \cite{seesaw}.
In the standard version of that mechanism
one introduces three gauge-singlet right-handed neutrinos $\nu_R$.
Let us define
\be
\nu^\prime_L \equiv C \bar \nu_R^T,
\ee
where $C$ is the charge-conjugation matrix.
Then,
the mass terms for the neutrinos are given by \cite{schechter}
\ba
\mathcal{L}_{\nu\,{\rm mass}}
&=& - \bar \nu_R M_D \nu_L - \bar \nu_L M_D^\dagger \nu_R
- \frac{1}{2}\, \bar \nu_R M_R C \bar \nu_R^T
+ \frac{1}{2}\, \nu_R^T C^{-1} M_R^\ast \nu_R
\label{D+M} \\
&=&
\frac{1}{2}
\left( \begin{array}{cc} \nu_L^T , & {\nu_L^\prime}^T
\end{array} \right) C^{-1}
\left( \begin{array}{cc} 0 & M_D^T \\ M_D & M_R
\end{array} \right)
\left( \begin{array}{c} \nu_L \\ \nu_L^\prime
\end{array} \right) + {\rm h.c.},
\label{mgkea}
\ea
where $M_R$ is a symmetric matrix.
If the eigenvalues of $M_R M_R^\ast$ are all much larger
than the eigenvalues of $M_D M_D^\dagger$,
then the approximate Majorana mass matrix for the light neutrinos
is given by
\be
{\cal M}_\nu = - M_D^T M_R^{-1} M_D,
\label{lightnu}
\ee
while the Majorana mass matrix for the heavy neutrinos
is approximately equal to $M_R$ \cite{fullseesaw}.
In the weak basis where the mass matrix $M_\ell$ of the charged leptons
is diagonal,
$M_\ell = {\rm diag} \left( m_e,  m_\mu, m_\tau \right)$,
one has
\be
U^T {\cal M}_\nu U = {\rm diag} \left( m_1, m_2, m_3 \right),
\label{U}
\ee
where $m_{1,2,3}$ are real non-negative
and $U$ is once again the lepton mixing matrix.

Since leptogenesis needs $CP$ violation, 
an intriguing question is whether
there is a connection
between the $CP$ violation at the seesaw scale and the one at low energies.
In the most general case,
the answer to this question is negative \cite{casas,branco1}.
However,
it is easy to find scenarios where such a connection exists---see,
for instance, 
\cite{branco2,pascoli};
in minimal scenarios only two heavy Majorana neutrinos are required
\cite{endoh,raidal,barger,felipe}. 
Inspired by grand unified theories,
it is also quite common in studies of leptogenesis
to assume hierarchies in the neutrino sector---see
\cite{falcone,akhmedov,velasco,rodejohann} and references therein.  
In particular,
one may assume that the Dirac neutrino mass matrix $M_D$
is strongly hierarchical,
i.e.\ that the eigenvalues $\left| a \right|^2$,
$\left| b \right|^2$,
$\left| c \right|^2$ of $M_D M_D^\dagger$ satisfy
$\left| a \right| \ll \left| b \right| \ll \left| c \right|$,
and subsequently reconstruct the masses of the heavy Majorana neutrinos
from the low energy data \cite{falcone,akhmedov}.
The assumption of a hierarchy in $M_D$
is justified by the relationship,
existing in some grand unified theories,
between $M_D$ and the up-type-quark mass matrix,
and by the fact that the latter matrix
is known to be strongly hierarchical.

In this paper we take a different departure and start from 
the fact that atmospheric neutrino mixing is maximal (or nearly maximal),
which suggests an alternative possibility \cite{Z2,D4,review-models}.
Since experimentally
$\left| U_{\mu 3} \right| \simeq \left| U_{\tau 3} \right|$,
there may exist in nature a $\mu$--$\tau$ interchange symmetry.
We know that $m_\mu \neq m_\tau$,
hence the $\mu$--$\tau$ interchange symmetry
must be broken in the charged-lepton sector,
but it may be kept intact in the neutrino sector.
This can be achieved through the introduction of three Higgs doublets,
one of them ($\phi_1$) with Yukawa couplings to the neutrino singlets
and to the charged-lepton singlet $e_R$,  
and the other two ($\phi_2$ and $\phi_3$) with Yukawa couplings
only to the charged-lepton singlets $\mu_R$ and $\tau_R$.
Under the interchange $\mu \leftrightarrow \tau$ the doublet $\phi_2$
remains invariant while $\phi_3$ changes sign;
this leads to $m_\mu \neq m_\tau$.
On the other hand,
the neutrino Dirac mass matrix is twofold degenerate
because $\phi_1$ is invariant under $\mu \leftrightarrow \tau$:
\be
M_D = {\rm diag} \left( a, b, b \right).
\label{uhyto}
\ee
A crucial feature of these models is the existence
of some other symmetry---either the continuous lepton-number symmetries
\cite{Z2} or a discrete symmetry \cite{D4}---which forces $M_D$
and the charged-lepton mass matrix $M_\ell$
to be \emph{simultaneously} diagonal.
These other symmetries are allowed to be \emph{softly} broken,
hence the right-handed-neutrino Majorana mass matrix is non-diagonal
and has the form
\be
M_R = \left( \begin{array}{ccc}
m & n & n \\ n & p & q \\ n & q & p
\end{array} \right)
\label{irtyx}
\ee
due to the $\mu$--$\tau$ interchange symmetry.
It is this matrix $M_R$ which produces lepton mixing.

The neutrino sectors of the
$\mathbbm{Z}_2$ model of \cite{Z2}
and of the $D_4$ model of \cite{D4}
are both characterized by equations (\ref{uhyto}) and (\ref{irtyx});
the $D_4$ model is more constrained than the $\mathbbm{Z}_2$ model
since it has $q = 0$ in $M_R$. 
Note that the $CP$-violating phase analogous to the CKM phase is
absent from the models under discussion, because $U_{e3} = 0$.
This follows easily from equations~(\ref{uhyto})
and (\ref{irtyx}), since 
\be
{\cal M}_\nu = \left( \begin{array}{ccc}
x & y & y \\ y & z & w \\ y & w & z
\end{array} \right) 
\label{Mnu}
\ee
in the basis in which $M_\ell$ is diagonal. 
Thus the only sources of $CP$ violation in the leptonic sector
are the two physical Majorana phases in $U$.

The purpose of this paper consists in analyzing leptogenesis
in the models of \cite{Z2,D4}. In particular, we will show that they
have the following properties:
\begin{enumerate}
\item Leptogenesis is a viable scenario. 
\item Correctly reproducing $\eta_B$
constrains the 
spectra of both the light and the heavy neutrinos. 
\item Only one of the two Majorana phases is responsible
for leptogenesis, and that phase
is also the only one which appears in the effective mass 
$\left| \langle m \rangle \right|$
for neutrinoless $\beta\beta$ decay. 
\end{enumerate}

In section~\ref{baryogenesis} we review the 
computation of $\eta_B$ from the knowledge of $M_D$ and $M_R$,
with emphasis on the three-Higgs-doublet structure of our models.
We proceed in section~\ref{analytical}
to derive the relevant analytic formulae
for the diagonalization of $M_R$ and $\mathcal{M}_\nu$,
in order to calculate $\eta_B$.
We apply those formulae in section~\ref{numerical}
to study the variation of $\eta_B$ with the parameters of the models.
In section~\ref{conclusions} we draw our conclusions.
An appendix contains calculational details
related to section~\ref{analytical}.

\section{Baryogenesis from leptogenesis}
\label{baryogenesis}

The ``natural'' basis for our models
is given by diagonal matrices $M_D$ and
$M_\ell$ while $M_R$ is non-diagonal---see 
equations~(\ref{uhyto}) and (\ref{irtyx}).
However,
the basis in which the leptogenesis formalism is established
is the one where $M_\ell$ and the
mass matrix of the right-handed neutrino singlets are diagonal;
the latter matrix is then 
\be
\hat M_R \equiv {\rm diag} \left( M_1, M_2, M_3 \right),
\label{Vint}
\ee
with real non-negative diagonal elements.
Defining a unitary matrix $V$ by
\be
V^T M_R V = \hat M_R,
\label{V}
\ee
we find, using equation~(\ref{D+M}),
\be\label{MDlepto}
M^\prime_D = V^T M_D 
\ee
for the neutrino Dirac mass matrix
in the leptogenesis basis.
For the actual calculation of $\eta_B$ one needs
\be\label{R}
R \equiv M^\prime_D {M^\prime_D}^\dagger = V^T M_D M_D^\dagger V^\ast.
\ee

We assume,
for the masses of the heavy Majorana neutrinos $N_{1,2,3}$,
that $M_1 \ll M_{2,3}$.
Then,
the $CP$ asymmetry $\epsilon_1$ produced in the decay of $N_1$
(the heavy neutrino with mass $M_1$)
is \cite{lepto-reviews,LB}
\be
\epsilon_1
= \frac{1}{8 \pi \left| v_1 \right|^2 R_{11}}
\sum_{j=2}^3\, f \left( \frac{M_j^2}{M_1^2} \right)
{\rm Im} \left[ \left( R_{1j} \right)^2 \right],
\label{formMD}
\ee
where $v_1$ denotes the vacuum expectation value (VEV) of $\phi_1^0$.
The function $f$ is given by
\be
f \left( t \right) = \sqrt{t} \left[ \frac{2 - t}{1 - t}
+ \left( 1 + t \right) \ln{\frac{t}{1 + t}} \right].
\ee
For $t \gg 1$, 
we have 
\be
f \left( t \right) =
- \frac{3}{2}\, t^{-1/2}
- \frac{5}{6}\, t^{-3/2}
- \frac{13}{12}\, t^{-5/2}
- \frac{19}{20}\, t^{-7/2}
- \cdots.
\ee
Thus,
$f \left( t \right)$ is negative.

The leptonic asymmetry produced through the decay of $N_1$ is
written as \cite{lepto-reviews,takanishi,pilaftsis}
\be
Y_L \equiv \frac{n_L - \bar n_L}{s} = \frac{\epsilon_1 \kappa_1}{g_{*1}},
\label{Y}
\ee
where $n_L$ is the lepton density,
$\bar n_L$ is the anti-lepton density,
$s$ is the entropy density,
$\kappa_1$ is the dilution factor
for the $CP$ asymmetry $\epsilon_1$
and $g_{*1}$ is the effective number of degrees of freedom
at the temperature $T = M_1$. 
The effective number of degrees of freedom is given by
(see for instance \cite{roos})
\be
g_* = \sum_{j = \mathrm{boson}} \! g_j +
\frac{7}{8} \sum_{k = \mathrm{fermion}} \! g_k.
\ee
In the $SU(2) \times U(1)$ gauge theory with three Higgs doublets
and supplemented by the seesaw mechanism one has 
\be\label{g1}
g_{*1} = \left[ 28 + \frac{7}{8} \times 90 \right]_\mathrm{SM} + 
8 + \frac{7}{8} \times 2
= 116.5,
\ee
where the terms within the brackets are the Standard Model
contributions and the last two terms in the sum
take into account the two additional Higgs doublets
and the lightest heavy neutrino $N_1$,
respectively.

The baryon asymmetry $Y_B$
produced through the sphaleron transmutation of $Y_L$,
while the quantum number $B-L$ remains conserved,
is given by \cite{harvey}
\be
Y_B = \frac{\omega}{\omega-1}\, Y_L \quad \mbox{with} \quad
\omega = \frac{8 N_F + 4 N_H}{22 N_F + 13 N_H},
\label{mgjdp}
\ee
where $N_F = 3$ is the number of fermion families
and $N_H$ is the number of Higgs doublets.
This relation derives from the thermal equilibrium of sphalerons 
for $10^2\, {\rm GeV} \lesssim T \lesssim 10^{12}\, {\rm GeV}$
\cite{lepto-reviews,pilaftsis}. 
Note that we must impose the condition $M_1 \lesssim 10^{12}\, {\rm GeV}$,
otherwise $Y_L$ could be erased before it transmutes into $Y_B$.
From equation~(\ref{mgjdp}),
$\omega = 12/35$ in three-Higgs-doublet models.

Next we discuss the relation between $Y_B$ and $\eta_B$,
where the latter quantity is the ratio baryon-number density
over photon density.
Note that this is the \emph{present} ratio,
and $\eta_B$ has last changed at the time of  $e^+ e^-$ annihilation,
at which time the photon density (temperature)
has increased relative to the neutrino density (temperature).
On the other hand,
$Y_B$ did \emph{not} change since its generation---the baryon number
per comoving volume
and the entropy per comoving volume remain constant.
Thus we have
\be
\eta_B = \left. \frac{s}{n_\gamma} \right|_0 Y_B,
\ee
where $n_\gamma$ is the photon density.
The index $0$ denotes \emph{present} time.
We know that \cite{roos}
\be
s = \frac{2 \pi^2}{45}\, g_{*0} T^3 
\quad \mbox{and} \quad
n_\gamma = \frac{2}{\pi^2}\, \zeta \left( 3 \right) T^3,
\ee
where $T$ is the photon temperature
and $\zeta \left( 3 \right) \simeq 1.20206$,
$\zeta$ being the zeta function.
At present only photons and light neutrinos are relevant for $s$,
since all other particles have annihilated apart from tiny remnants---actually,
we are just discussing the baryon remnant.
Thus
\be
g_{*0} = 2 + \frac{7}{8} \times 6 \times \frac{4}{11}
= \frac{43}{11},
\ee
where we have taken into account that
the neutrino temperature after $e^+ e^-$ annihilation
is a factor of $\left( 4/11 \right)^{1/3}$ lower than the
photon temperature $T$.
One thus obtains \cite{barger,buchmueller} 
\be\label{etaB}
\left. \frac{s}{n_\gamma} \right|_0 =
\frac{\pi^4}{\zeta(3)}\, \frac{43}{495}
= 7.0394 \quad \mbox{and} \quad
\eta_B = 7.0394\, Y_B \simeq -3.15 \times 10^{-2}\, \kappa_1 \epsilon_1.
\ee
We have used the values for three-Higgs-doublet models.

We now turn to the dilution factor,
which is approximately given by \cite{takanishi,kolb,notari} 
\be
\kappa_1 \simeq \frac{0.3}{K_1 \left( \ln K_1 \right)^{3/5}},
\label{kappa}
\ee
where
\be
K_1 \equiv \frac{\Gamma_1}{H_1}.
\ee
In this equation,
$\Gamma_1$ is the decay width of $N_1$,
given at tree level by \label{lepto-reviews}
\be
\Gamma_1 = \frac{R_{11} M_1}{8 \pi \left| v_1 \right|^2},
\ee
and $H_1$ is the Hubble constant at temperature $T = M_1$,
\be
H_1 = 1.66\, \sqrt{g_{*1}}\, \frac{M_1^2}{M_{\rm Planck}},
\ee
where $M_{\rm Planck} = 1.221 \times 10^{19}\, \mathrm{GeV}$.
Thus,
\be
K_1 = \frac{M_{\rm Planck}\, R_{11}}
{1.66\, \sqrt{g_{*1}}\, 8 \pi \left| v_1 \right|^2 M_1}.
\label{urtwn}
\ee
Equation~(\ref{kappa}) holds
for $10 \lesssim K_1 \lesssim 10^6$ \cite{takanishi,kolb,notari}.
Numerically one obtains
\be
K_1 \simeq \frac{895.6}{1\, {\rm eV}}\, \frac{R_{11}}{M_1}
\left( \frac{174\, \mathrm{GeV}}{\left| v_1 \right|} \right)^2.
\label{numericalK1}
\ee
For $g_{*1}$ we have used the value of equation~(\ref{g1}).

With equations~(\ref{formMD}), (\ref{etaB}), 
(\ref{kappa}) and (\ref{urtwn}) one obtains,
after a numerical evaluation,
\be
\eta_B \simeq -1.39 \times 10^{-9}\,
\frac{1}{\left( \ln{K_1} \right)^{3/5}}
\sum_{j=2}^3\, f \left( \frac{M_j^2}{M_1^2} \right)
\frac{{\rm Im} \left[ \left( R_{1j} \right)^2 \right]}
{\left( R_{11} \right)^2}\,
\frac{M_1}{10^{11}\, {\rm GeV}}.
\label{finaletaB}
\ee
This equation,
which we shall use in conjunction with equation~(\ref{numericalK1}),
clearly indicates the desired order of magnitude of $M_1$.
Equation~(\ref{finaletaB}) 
shows that $\eta_B$ is 
approximately independent of $\left| v_1 \right|$.
Only the logarithm of $K_1$ in the denominator makes
$\eta_B$ dependent on $\left| v_1 \right|$.
Also notice that the minus sign in equation~(\ref{finaletaB}) cancels
with the negative sign of $f \left( t \right)$.

\section{The baryon asymmetry in our models}
\label{analytical}

It remains to calculate the quantities 
${\rm Im} \left[ \left( R_{1j} \right)^2 \right]$ ($j = 2,3$) 
and $R_{11}$ in our models. 
We will see that the $\mathbbm{Z}_2$ model allows a full analytical
calculation of these quantities and, according to the discussion at
the end of section \ref{introduction}, these quantities in the $D_4$
model are special cases of those in the $\mathbbm{Z}_2$ model.

We remind the reader that,
in our models,
the effective light-neutrino Majorana mass matrix $\mathcal{M}_\nu$
is as in equation~(\ref{Mnu}),
while the right-handed-neutrino Majorana mass matrix $M_R$
has the same form as $\mathcal{M}_\nu$ and is in equation~(\ref{irtyx}).
The matrix $\mathcal{M}_\nu$ is diagonalized by $U$,
see equation~(\ref{U}),
while $M_R$ is diagonalized by $V$,
see equations~(\ref{Vint}) and~(\ref{V}).
The analogies between $\mathcal{M}_\nu$ and $M_R$
and between $U$ and $V$
are used repeatedly throughout this paper.

It can be shown that,
because $\mathcal{M}_\nu$ is of the specific form
in equation~(\ref{Mnu}),
i.e.\ with $\left( \mathcal{M}_\nu \right)_{\mu \mu}
= \left( \mathcal{M}_\nu \right)_{\tau \tau}$
and $\left( \mathcal{M}_\nu \right)_{e \mu}
= \left( \mathcal{M}_\nu \right)_{e \tau}$,
the matrix $U$ can be parametrized as
\be
U = {\rm diag} \left( 1, e^{i \alpha}, e^{i \alpha} \right)
\left( \begin{array}{ccc}
-c & s & 0 \\
s/\sqrt{2} & c/\sqrt{2} & 1/\sqrt{2} \\
s/\sqrt{2} & c/\sqrt{2} & -1/\sqrt{2}
\end{array} \right)
{\rm diag} \left( e^{i \beta_1}, e^{i \beta_2}, e^{i \beta_3} \right),
\label{jgidw}
\ee
with $c \equiv \cos{\theta}$
and $s \equiv \sin{\theta}$.\footnote{Equation~(\ref{jgidw})
is clearly not the most general parametrization for a unitary matrix,
rather it is a consequence of the specific form of $\mathcal{M}_\nu$
in equation~(\ref{Mnu}).}
We assume,
without loss of generality,
that $\theta$ belongs to the first quadrant while $m_2 > m_1$.
The phase $\alpha$ is unphysical;
the only physical phases are the differences
\be\label{Delta}
\Delta \equiv 2 (\beta_1 - \beta_2)
\ee
and $2 (\beta_1 - \beta_3)$.
The matrix $\mathcal{M}_\nu$ has six physical parameters:
the moduli of $x$,
$y$,
$z$ and $w$ and the phases
of $z w^\ast$ and $y^2 x^\ast z^\ast$.\footnote{The physical phases
in $\mathcal{M}_\nu$ are the ones of its Jarlskog invariants.
For instance,
$\arg{\left( z w^\ast \right)}$ is the phase of
$\left( \mathcal{M}_\nu \right)_{\mu \mu}
\left( \mathcal{M}_\nu \right)_{e \tau}
\left( \mathcal{M}_\nu \right)_{\mu \tau}^\ast
\left( \mathcal{M}_\nu \right)_{e \mu}^\ast$,
and $\arg{\left( y^2 x^\ast z^\ast \right)}$ is the phase of
$\left( \mathcal{M}_\nu \right)_{e \mu} \left( \mathcal{M}_\nu \right)_{\mu e}
\left( \mathcal{M}_\nu \right)_{ee}^\ast
\left( \mathcal{M}_\nu \right)_{\mu \mu}^\ast$.}
It is better to use as physical parameters the moduli of $z + w$
and $z - w$ instead of the moduli of $z$ and $w$,
and the phases of $\left( z - w \right) \left( z + w \right)^\ast$
and $y^2 x^\ast \left( z + w \right)^\ast$
instead of the phases of $z w^\ast$ and $y^2 x^\ast z^\ast$.
The six parameters $\left| x \right|$,
$\left| y \right|$,
$\left| z + w \right|$,
$\left| z - w \right|$,
$\arg \left[ y^2 x^\ast \left( z + w \right)^\ast \right]$
and $\arg \left[ \left( z - w \right) \left( z + w \right)^\ast \right]$ 
correspond to the six observables $m_1$,
$m_2$,
$m_3$,
$\theta$,
$\Delta$ and $2 (\beta_1 - \beta_3)$.

Since the matrices $M_R$ and $\mathcal{M}_\nu$ have the same form,
the matrix $V$ which diagonalizes $M_R$
has the same form as the matrix $U$ which diagonalizes $\mathcal{M}_\nu$:
\be
V = {\rm diag} \left( 1, e^{i \chi}, e^{i \chi} \right)
\left( \begin{array}{ccc}
-c^\prime & s^\prime & 0 \\
s^\prime/\sqrt{2} & c^\prime/\sqrt{2} & 1/\sqrt{2} \\
s^\prime/\sqrt{2} & c^\prime/\sqrt{2} & -1/\sqrt{2}
\end{array} \right)
{\rm diag} \left( e^{i \gamma_1}, e^{i \gamma_2}, e^{i \gamma_3} \right),
\label{jgieq}
\ee
with $c^\prime \equiv \cos{\theta^\prime}$
and $s^\prime \equiv \sin{\theta^\prime}$.
We assume,
without loss of generality,
that $\theta^\prime$ belongs to the first quadrant while $M_2 > M_1$.
Like in the previous paragraph,
there are six observables originating in $M_R$:
$M_1$,
$M_2$,
$M_3$,
$\theta^\prime$, 
$2(\gamma_1 - \gamma_2)$ and $2(\gamma_1 - \gamma_3)$.

We proceed to the calculation of $R$---see equation~(\ref{R}).  
Using equations~(\ref{uhyto}) and (\ref{jgieq}),
we obtain
\be
R = V^T M_D M_D^\dagger V^\ast
= \left( \begin{array}{ccc}
\left| a \right|^2 {c^\prime}^2 + \left| b \right|^2 {s^\prime}^2
&
c^\prime s^\prime \left( \left| b \right|^2 - \left| a \right|^2 \right)
e^{i \left( \gamma_1 - \gamma_2 \right)}
&
0 \\
c^\prime s^\prime \left( \left| b \right|^2 - \left| a \right|^2 \right)
e^{i \left( \gamma_2 - \gamma_1 \right)}
&
\left| a \right|^2 {s^\prime}^2 + \left| b \right|^2 {c^\prime}^2
&
0 \\
0 & 0 & \left| b \right|^2
\end{array} \right).
\label{mgowx}
\ee
The fact that 
\be
R_{13} = R_{23} = 0 
\ee
implies that
in this model the third heavy neutrino has no bearing
on leptogenesis---even if its mass $M_3$ happens to be lower
than $M_1$ and $M_2$,
the masses of the other two heavy neutrinos.

The matrices $\mathcal{M}_\nu$ and $M_R$ are related
through equation~(\ref{lightnu}),
with $M_D$ given by equation~(\ref{uhyto}).
That relation contains 
only two extra parameters:
$\left| a \right|$ and $\left| b \right|$.
This means that,
out of the six observables originating in $M_R$,
only two can be considered as independent of the six observables
originating in $\mathcal{M}_\nu$.
We shall select $M_{1,2}$,
together with the observables originating in $\mathcal{M}_\nu$,
as the basic observables of the theory,
and express the four remaining ones in terms of these.
It will turn out that $2 \left( \beta_1 - \beta_3 \right)$
and $m_3$ play no role in the computation of $\eta_B$;
the basic observables that we need for that computation are
\be
m_{1,2}, \; M_{1,2}, \; \theta \; \mbox{and} \; \Delta.
\label{physpar}
\ee

With the aim of using the experimental information on 
$\Delta m^2_\odot$ and $\Delta m^2_\mathrm{atm}$,
we express
\be\label{m23}
m_2 = \sqrt{m_1^2 + \Delta m^2_\odot} 
\quad \mbox{and} \quad
m_3 = \sqrt{m_1^2 + \sigma \Delta m^2_\mathrm{atm}}
\quad (\sigma = \pm 1)
\ee
as functions of $m_1$.
For $m_3$ there is a twofold ambiguity
stemming from the two possible neutrino spectra:
normal spectrum $m_1 < m_2 < m_3$ ($\sigma = +1$) 
and inverted spectrum $m_3 < m_1 < m_2$ ($\sigma = -1$).
The inverted spectrum is only allowed for 
$m_1 \geq \sqrt{\Delta m^2_\mathrm{atm}}$;
the experimental result
for the atmospheric mass-squared difference---see
equation~(\ref{dm2atm})---then requires $m_1 \gtrsim 0.04\, \mathrm{eV}$,
which is rather large.

As shown in the Appendix,
the parameters $\left| x \right|$,
$\left| y \right|$ and $\left| z + w \right|$ are given
in terms of the observables in equation~(\ref{physpar}) by
\ba
\left| x \right| &=& \left| c^2 m_1 + s^2 m_2 e^{i \Delta} \right|,
\label{xabs} \\
\left| y \right| &=& \frac{c s}{\sqrt{2}}
\left| m_1 - m_2 e^{i \Delta} \right|,
\label{yabs} \\
\left| z + w \right| &=& \left| s^2 m_1 + c^2 m_2 e^{i \Delta} \right|.
\label{z+w}
\ea
We may then compute
\ba
B &=& 4 m_1 m_2 M_1 M_2 \left| y \right|^2
- m_1^2 m_2^2 \left( M_1^2 + M_2^2 \right), \label{Beq} \\
C &=& m_1^2 m_2^2 M_1^2 M_2^2
\left| x \right|^2 \left| z + w \right|^2. \label{Ceq}
\ea
These are the coefficients of the equation
\be
\left| x \right|^4 \left| b \right|^8
+ B \left| x \right|^2 \left| b \right|^4 + C = 0,
\label{afhro}
\ee
which is derived in the Appendix
and allows one to compute the parameter $\left| b \right|$
as a function of the observables in equation~(\ref{physpar}).
Indeed,
\be
\left| b \right|^4
= \frac{- B \pm \sqrt{B^2 - 4 C}}{2 \left| x \right|^2}.
\label{bbbbb}
\ee
Moreover,
the parameter $\left| a \right|$ is given by---see equation~(\ref{nrtso})
\be
\left| a \right|^4
= \frac{- B \mp \sqrt{B^2 - 4 C}}{2 \left| z + w \right|^2},
\label{aaaaa}
\ee
and the mixing angle $\theta^\prime$ is given,
as a function of the observables in equation~(\ref{physpar}),
by---see equation~(\ref{iuwem})
\be
{c^\prime}^2 - {s^\prime}^2
= \frac{\pm \sqrt{B^2 - 4 C}}{m_1^2 m_2^2 \left( M_2^2 - M_1^2 \right)}.
\label{prime}
\ee
Thus,
using $m_{1,2}$,
$M_{1,2}$,
$\theta$ and $\Delta$ as input,
we are able to compute,
with a twofold ambiguity,
$\left| b \right|$,
$\left| a \right|$ and $\theta^\prime$.
The twofold ambiguity corresponds to the interchanges
$\left| m \right| \leftrightarrow \left| p + q \right|$
and ${c^\prime}^2 \leftrightarrow {s^\prime}^2$,
as is made clear in the Appendix.

One must check that the condition $B^2 - 4 C \ge 0$ is respected.
This condition leads to a lower bound on $m_1$;
in the limit $M_2 \gg M_1$ one finds approximately
\be
\frac{m_1}{m_2} \gtrsim
\frac{M_1}{M_2}\, \sin^2{2 \theta} 
\quad \mbox{or} \quad 
m_1 \gtrsim \sqrt{\Delta m^2_\odot}\,
\frac{M_1}{M_2}\, \sin^2{2 \theta}.
\label{lowerbound}
\ee
On the other hand,
it turns out that,
in general,
with $m_2$ given by the first equation~(\ref{m23}),
$\left| a \right|^2$ and $\left| b \right|^2$
increase with increasing $m_1$.
In our models,
$a$ and $b$ arise from the Yukawa Lagrangian
\be
{\cal L}_{\rm Y} =
\frac{\left( - \phi_1^0\, , \ \phi_1^+ \right)}{v_1} \left\{
a\, \bar \nu_{eR} \left( \begin{array}{c} \nu_{eL} \\ e_L \end{array} \right)
+ b \left[
\bar \nu_{\mu R} \left( \begin{array}{c} \nu_{\mu L} \\ \mu_L
\end{array} \right)
+ \bar \nu_{\tau R} \left( \begin{array}{c} \nu_{\tau L} \\ \tau_L
\end{array} \right) \right] \right\}
+ {\rm h.c.} - \cdots,
\label{qurns}
\ee
where $\cdots$ represents the Yukawa couplings
of the right-handed charged-lepton fields.
If we require that the Yukawa coupling constants $a / v_1$ and $b / v_1$
should at most be of order 1,
this implies a loose upper bound on $m_1$
(when $m_2^2 - m_1^2$ is kept fixed),
since $\left| a \right|$ and $\left| b \right|$ increase with $m_1$.
For $M_2 \gg M_1$
and using the solution $|b|^2 > |a|^2$,
i.e.\ the upper signs in equations~(\ref{bbbbb}) and (\ref{aaaaa}),
we find the approximate expressions 
\ba
\left| a \right|^2
&\simeq& m_1 M_1 \left[ 1 - \sin^2{\left( \Delta/2 \right)}
\sin^2{2 \theta} \right]^{1/2}, \\
\left| b \right|^2
&\simeq& m_1 M_2 \left[ 1 - \sin^2{\left( \Delta/2 \right)}
\sin^2{2 \theta} \right]^{-1/2},
\ea
where we have used $m_2 \simeq m_1$,
valid for $m_1 \gg \sqrt{\Delta m^2_\odot}$. 
These equations may be used to compute
the approximate upper bound on $m_1$.
We see that $\left| b \right|^2 \sim m_1 M_2$.
Therefore,
requiring
$\left| b \right|^2 \left/ \left| v_1 \right|^2 \right. \lesssim 1$
leads to
\be
m_1 \lesssim \frac{\left| v_1 \right|^2}{M_2}.
\label{maximum}
\ee
For instance,
for $\left| v_1 \right| = 10\, {\rm GeV}$
and $M_2 = 10^{13}\, {\rm GeV}$,
equation~(\ref{maximum}) yields $m_1 \lesssim 0.01\, {\rm eV}$.
If we choose $\left| v_1 \right| = 50\, {\rm GeV}$
and $M_2 = 2.5 \times 10^{12}\, {\rm GeV}$, we have 
$m_1 \lesssim 1\, {\rm eV}$.

An expression for ${\rm Im} \left[ \left( R_{12} \right)^2 \right]$
in terms of our basic observables is derived in the Appendix,
see equation~(\ref{imimim}).
Using  equation~(\ref{finaletaB}),
the main result of this section is 
\be
\eta_B \simeq -1.39 \times 10^{-9}\,
\frac{1}{\left( \ln{K_1} \right)^{3/5}}\,
f \left( \frac{M_2^2}{M_1^2} \right)
\left( \left| b \right|^2 - \left| a \right|^2 \right)^2
\frac{M_1^2 M_2 \left( m_2^2 - m_1^2 \right) c^2 s^2 \sin{\Delta}}
{m_1 m_2 \left( M_2^2 - M_1^2 \right) \left( R_{11} \right)^2
\left( 10^{11}\, {\rm GeV} \right)}.
\label{main}
\ee
Here,
a convenient expression for $R_{11}$---see equation~(\ref{mgowx})---is
\be
R_{11} = \frac{1}{2} \left[\, |a|^2 + |b|^2 + 
(|a|^2 - |b|^2) \left ({c^\prime}^2 - {s^\prime}^2 \right) \right],
\label{R11}
\ee
where ${c^\prime}^2 - {s^\prime}^2$ is given by equation~(\ref{prime}).

Since $f \left( M_2^2 \left/ M_1^2 \right. \right)$ is negative,
we find from equation~(\ref{main}) that $\sin{\Delta} > 0$.
Notice the important point that the violation of $CP$
responsible for the generation of a non-zero $\eta_B$
all comes from the Majorana phase $\Delta$.
This is the same Majorana phase entering the matrix element
for neutrinoless double beta decay,
\be
\left| \langle m \rangle \right| = 
\left| \left( \mathcal{M}_\nu \right)_{ee} \right|
= \left| x \right|, 
\label{betabetamass}
\ee
which is given by equation~(\ref{xabs}).
Thus,
in these models neutrinoless double beta decay and leptogenesis
depend on the same Majorana phase $\Delta$.

To conclude this section, we discuss 
the extra condition on $M_R$ in the $D_4$ model of \cite{D4}.
As mentioned before, in that model
$\left( M_R \right)_{\mu \tau} = q = 0$.
Since $M_D$ is diagonal,
this leads to $\left( \mathcal{M}_\nu^{-1} \right)_{\mu \tau} = 0$,
or \cite{D4}
\be
\frac{s^2}{m_1}\, e^{2 i \beta_1}
+ \frac{c^2}{m_2}\, e^{2 i \beta_2}
- \frac{1}{m_3}\, e^{2 i \beta_3} = 0.
\ee
The Majorana phase $2 (\beta_1 - \beta_3)$ is irrelevant for our purposes,
therefore the important constraint is
\be
\frac{1}{m_3} = \left| \frac{s^2}{m_1}\, e^{i \Delta} +
\frac{c^2}{m_2} \right|.
\ee
This forces $m_3$ to be larger than both $m_1$ and $m_2$
(normal spectrum).
The Majorana phase $\Delta$ becomes a function of $m_1$ through
\be
\cos{\Delta} = \frac{\left( m_1 m_2 \left/ m_3 \right. \right)^2
- c^4 m_1^2 - s^4 m_2^2}{2 c^2 s^2 m_1 m_2}
\ee
and through equations~(\ref{m23}) with $\sigma = +1$.
Thus,
the $D_4$ model of \cite{D4}
has one degree of freedom less
than the $\mathbbm{Z}_2$ model of \cite{Z2}.

\section{Numerical results}
\label{numerical}

In this section we shall always use
\ba
\theta &=& 33^\circ, \label{valuetheta} \\
\Delta m^2_\odot &=& 7.1 \times 10^{-5}\, {\rm eV}^2, \label{valuem}
\ea
the best-fit values of \cite{SNO}.
Then the observables in our set~(\ref{physpar})
which are still free to be chosen are
$m_1$, $M_{1,2}$ and $\Delta$;
we furthermore have to choose
$|v_1|$---see equations~(\ref{numericalK1}) and (\ref{main}). 
The mass $m_2$ is fixed via equations~(\ref{m23}) and~(\ref{valuem}). 

From this input one obtains $\left| x \right|$,
$\left| y \right|$
and $\left| z + w \right|$ using equations~(\ref{xabs})--(\ref{z+w}).
One then computes $\left| a \right|$,
$\left| b \right|$ and $\theta^\prime$,
with a twofold ambiguity,
from equations~(\ref{bbbbb})--(\ref{prime}).
Thereafter,
$R_{11}$ is found in equation~(\ref{R11})
and $K_1$ is given by equation~(\ref{numericalK1}).

As for the twofold ambiguity
in the computation of $\left| a \right|$ and $\left| b \right|$,
we use the upper signs in equations~(\ref{bbbbb})--(\ref{prime}).
Numerically,
the choice of the lower signs
yields a smaller $\eta_B$.

The VEV $\left| v_1 \right|$
must be smaller than $174\, {\rm GeV}$,
in order that there is also room for the other two VEVs:
$\sum_{j=1}^3 \left| v_j \right|^2 = (174\, \mathrm{GeV})^2$.
Since in the models of \cite{Z2,D4}
both the neutrino masses and the electron mass
originate in Yukawa couplings to $\phi_1$,
while the $\mu$ and $\tau$ masses originate in Yukawa couplings
to $\phi_2$ and $\phi_3$,  
the smallness of the electron and neutrino masses
suggests that $v_1$ should be relatively small.

\begin{figure}[t]
\begin{center}
\epsfig{file=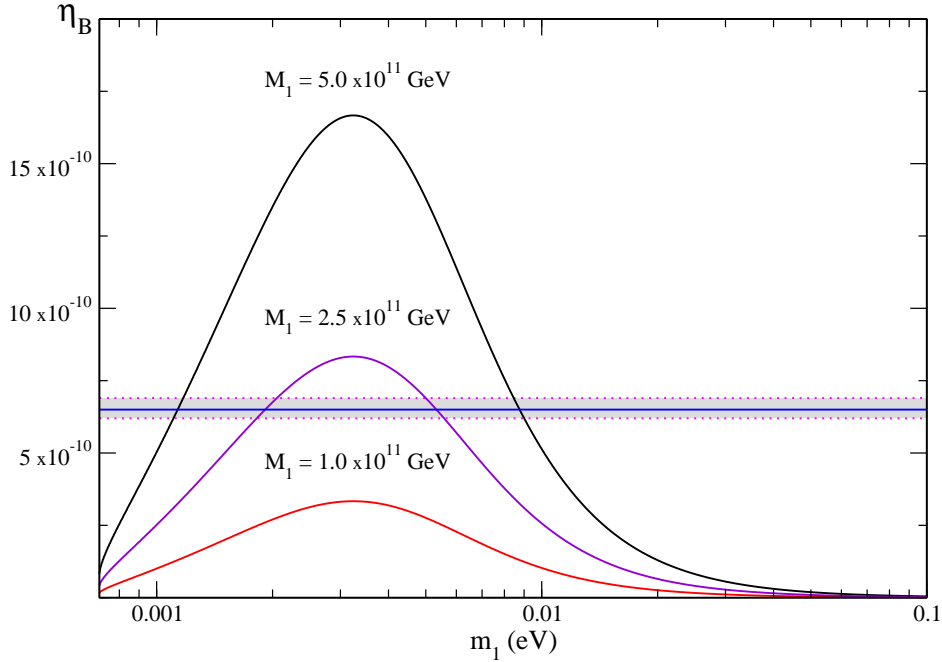,width=11cm,angle=-90}
\end{center}
\caption{$\eta_B$ as function of $m_1$
for three different values of $M_1$.
In producing this figure we have chosen $M_2/M_1 = 10$, 
$\left| v_1 \right| = 50\, \mathrm{GeV}$ and $\Delta = 90^\circ$.
The lowest allowed $m_1$ for this value $M_2/ M_1 = 10$
is $m_1 = 0.71 \times 10^{-3}\,  \mathrm{eV}$.
The horizontal lines indicate the experimental result for $\eta_B$,
as given in equation~(\ref{etaBexp}). \label{etavsm1}}
\end{figure}
In figure~\ref{etavsm1} we have plotted $\eta_B$ versus $m_1$ for three
different
values of $M_1$.
We read off from that figure that
$M_1$ must be larger than $10^{11}\, \mathrm{GeV}$ in order to
reproduce the experimental value of $\eta_B$ in equation~(\ref{etaBexp}).
Since a successful leptogenesis requires 
$M_1 < 10^{12}\, \mathrm{GeV}$ \cite{lepto-reviews,pilaftsis},
the order of magnitude of $M_1$
becomes quite constrained. 
We furthermore see that
small values of $m_1$ are preferred; for large values of $m_1$,
$\eta_B$ becomes too small.
Actually from figure~\ref{etavsm1}
we gather that $m_1$ cannot exceed $0.02\, \mathrm{eV}$.
This rules out the inverted neutrino spectrum
if we want to accommodate leptogenesis in our model.
As a function of $m_1$,
the maximum of the theoretical expression~(\ref{main})
for $\eta_B$ is roughly at
$m_1 = 3 \times 10^{-3}\, \mathrm{eV}$.
If we consider $\eta_B$
as a function of the $CP$-violating Majorana phase $\Delta$,
a numerical study shows
that the maximum of $\eta_B$ is attained for $\Delta$ close to $100^\circ$.
As a function of $\left| v_1 \right|$,
$\eta_B$ increases by less than a factor of two when that VEV
goes from $10\, \mathrm{GeV}$ to $100\, \mathrm{GeV}$.

In order to understand
the dependence of $\eta_B$ of equation~(\ref{main})
on $M_{1,2}$,
it is useful to perform a scale transformation $M_{1,2} \to \lambda M_{1,2}$
where $\lambda$ is an arbitrary positive number.
From equations~(\ref{bbbbb}) and (\ref{aaaaa})
we see that $\left| a \right|^2$ and $\left| b \right|^2$
scale with one power of $\lambda$;
the same holds for $R_{12}$ and $R_{11}$.
It follows that $\eta_B$ also scales with one power of $\lambda$;
in other words,
$\eta_B$ is a homogeneous function of order one
in $M_{1,2}$.

\begin{figure}[t]
\begin{center}
\epsfig{file=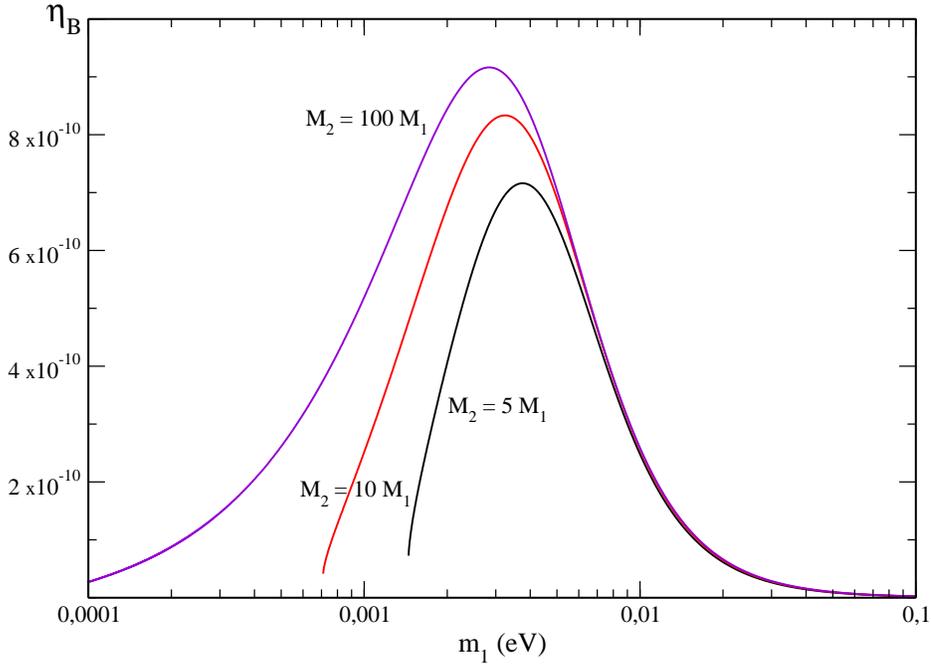,width=11cm,angle=-90}
\end{center}
\caption{$\eta_B$ as a function of $m_1$ for $M_1 = 2.5 \times
10^{11}\, \mathrm{GeV}$ and three different values of $M_2$. 
We have used $\left| v_1 \right| = 50\, \mathrm{GeV}$ and $\Delta =
90^\circ$. \label{etavsm1M2}}
\end{figure}
In figure~\ref{etavsm1M2}
we have plotted $\eta_B$ versus $m_1$
for $M_1 = 2.5 \times 10^{11}\, \mathrm{GeV}$
and different values of $M_2$.
This figure shows
that fixing $M_1$ and varying $M_2$ with $M_2\gg M_1$
does not drastically
alter $\eta_B$, except for very small values of $m_1$.
The lower bound on $m_1$ as a function of the ratio $M_2/M_1$
is also illustrated in figure~\ref{etavsm1M2}.

\section{Conclusions}
\label{conclusions}

\begin{figure}[t]
\begin{center}
\epsfig{file=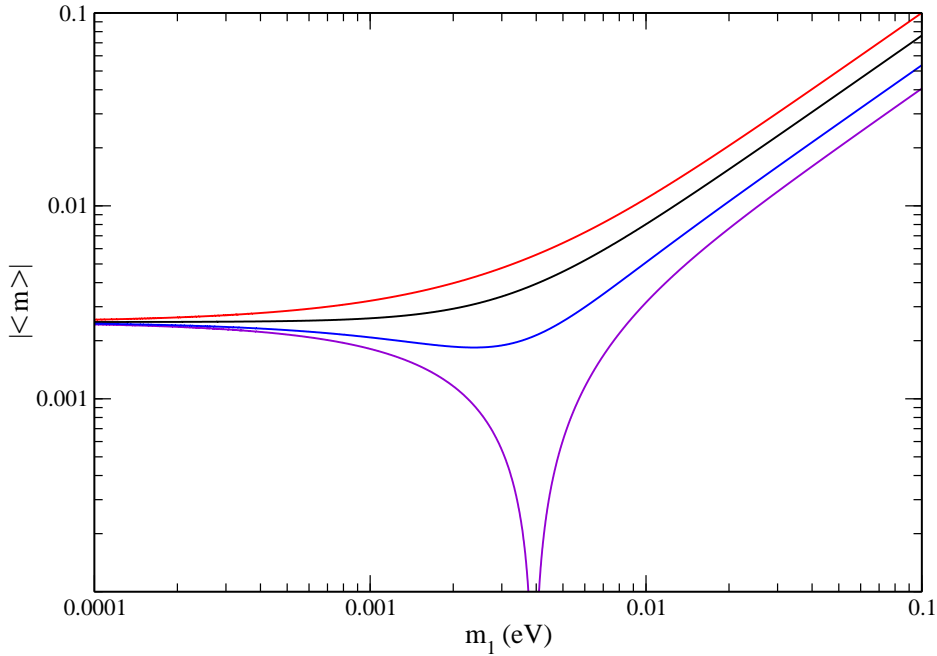,width=11cm,angle=-90}
\end{center}
\caption{The effective Majorana mass in neutrinoless $\beta\beta$
decay as a function of $m_1$.
Starting at the upper curve and descending to the lowest
of the four curves,
the values used for $\Delta$ are $0^\circ$,
$90^\circ$,
$135^\circ$ and $180^\circ$, respectively. \label{meff_fig}}
\end{figure}
In this paper we have
computed the baryon asymmetry of the universe
in the $\mathbbm{Z}_2$ model of \cite{Z2}
and in the $D_4$ model of \cite{D4}.
These models are characterized by
a neutrino Dirac mass matrix $M_D$
with two degenerate eigenvalues,
and by an interchange ($\mathbbm{Z}_2$)
symmetry between the $\mu$ and $\tau$ families.
Both models predict maximal
atmospheric neutrino mixing and $U_{e3} = 0$
as a consequence of their symmetries. 

We have shown that these models can easily accommodate
baryogenesis via leptogenesis
and reproduce the experimental value
of $\eta_B$,\footnote{On the other hand, as shown in \cite{akhmedov}, 
a mass hierarchy in $M_D$ requires finetuning of the masses in the 
heavy neutrino sector in order to reproduce $\eta_B$.} 
provided the mass of $N_1$,
the lightest heavy neutrino whose decays generate the relevant $CP$ asymmetry,
is in the range 
$10^{11}\, \mathrm{GeV} \lesssim M_1 \lesssim 10^{12}\, \mathrm{GeV}$.
Furthermore,
the mass $m_1$ of the lightest neutrino
has to be in the range 
$10^{-3}\, \mathrm{eV} \lesssim m_1 \lesssim 10^{-2}\, \mathrm{eV}$.
Thus, an inverted neutrino mass spectrum is practically excluded.

In figure~\ref{meff_fig} we have plotted the effective
mass $\left| \langle m \rangle \right|$ 
probed in neutrinoless $\beta\beta$ decay.
We see that, if our models are to
have successful leptogenesis,
then $\left| \langle m \rangle \right|$
is at most $0.01\, \mathrm{eV}$;
if the evidence for
neutrinoless $\beta\beta$ decay
of \cite{klapdor}, with $\left| \langle m \rangle \right| > 0.1\, \mathrm{eV}$
is confirmed, then
leptogenesis is not enough to generate an $\eta_B$ of the observed size.

We stress that the neutrino mass matrices of our models
allow an analytical calculation
of the $CP$ asymmetry $\epsilon_1$ of equation~(\ref{formMD}).
The results of this paper
for the $\mathbbm{Z}_2$ and $D_4$ models
may be valid in a wider context of general models with
a twofold
degenerate Dirac mass matrix $M_D$; the reason is that
the mass matrix of the light neutrinos must have the
form in equation~(\ref{Mnu}) if one 
assumes maximal atmospheric neutrino mixing
and $U_{e3} = 0$,
assumptions that, as experiment shows, cannot be far from true.

\vspace*{5mm}

\noindent \textbf{Acknowledgements:} 
We are grateful to Ricardo Gonz\'alez Felipe for helpful discussions.
The work of L.L.\ was supported
by the Portuguese \textit{Funda\c c\~ao para a Ci\^encia e a Tecnologia}
under the contract CFIF-Plurianual.

\newpage

\begin{appendix}
\setcounter{equation}{0}
\renewcommand{\theequation}{\Alph{section}\arabic{equation}}
\section{Algebraic details}

\paragraph{The matrices $M_R$ and $\mathcal{M}_\nu$}
Those matrices are given in equations~(\ref{irtyx}) and~(\ref{Mnu}),
and their relationship to each other
is expressed by equations~(\ref{lightnu}) and~(\ref{uhyto}).
We define
\ba
d &\equiv& 2 n^2 - m \left( p + q \right),
\label{mfkgh} \\
f &\equiv& 2 y^2 - x \left( z + w \right),
\label{mfjdp}
\ea
so that
\be
\det{M_R} = \left( q - p \right) d 
\quad \mbox{and} \quad
\det{\mathcal{M}_\nu} = \left( w - z \right) f.
\label{det}
\ee
Then,
by explicitly inverting $M_R$ and by using equation~(\ref{lightnu}),
we find
\ba
x &=& a^2\, \frac{p + q}{d}, \label{x} \\
y &=& a b\, \frac{- n}{d}, \label{y} \\
z + w &=& b^2\, \frac{m}{d} \label{zpw}
\ea
and
\be
\left( z - w \right) \left( q - p \right) = b^2.
\label{zmw}
\ee
It is also useful to invert equation~(\ref{lightnu}): 
\be
M_R = - M_D {\cal M}_\nu^{-1} M_D^T.
\label{tyruq}
\ee
From this relation we compute 
\ba
m &=& a^2\, \frac{z + w}{f}, \label{m} \\
n &=& a b\, \frac{- y}{f}, \label{n} \\
p + q &=& b^2\, \frac{x}{f}. \label{ppq}
\ea
In addition,
we obtain equation~(\ref{zmw}) again.

\paragraph{The parameters $x$, $y$, $z$ and $w$}
From equations~(\ref{U}),
(\ref{Mnu}) and~(\ref{jgidw}) one may write
\be
\left( \begin{array}{ccc}
x & y e^{i \alpha} & y e^{i \alpha} \\
y e^{i \alpha} & z e^{2 i \alpha} & w e^{2 i \alpha} \\
y e^{i \alpha} & w e^{2 i \alpha} & z e^{2 i \alpha}
\end{array} \right)
=
\left( \begin{array}{ccc}
-c & s & 0 \\ 
s/\sqrt{2} & c/\sqrt{2} & 1/\sqrt{2} \\
s/\sqrt{2} & c/\sqrt{2} & -1/\sqrt{2}
\end{array} \right)
\,\hat \mu\, 
\left( \begin{array}{ccc}
-c & s/\sqrt{2} & s/\sqrt{2} \\ 
s & c/\sqrt{2} & c/\sqrt{2} \\
0 & 1/\sqrt{2} & -1/\sqrt{2}
\end{array} \right),
\ee
where
\be
\hat \mu = 
\mbox{diag} \left( m_1 e^{- 2 i \beta_1},\, m_2 e^{- 2 i \beta_2},\,
m_3 e^{- 2 i \beta_3} \right).
\ee
From this we obtain equations
giving the parameters of $\mathcal{M}_\nu$
as functions of the physical observables.
We find equations~(\ref{xabs})--(\ref{z+w}) and also
\ba
y^2 x^\ast \left( z + w \right)^\ast &=&
\frac{c^2 s^2}{2} \left( m_1 - m_2 e^{i \Delta} \right)^2
\left( c^2 m_1 + s^2 m_2 e^{- i \Delta} \right)
\left( s^2 m_1 + c^2 m_2 e^{- i \Delta} \right), \hspace*{10mm}
\label{y2xzpw} \label{phase1} \\
\left| z - w \right| &=& m_3,
\label{z-w} \\
\left( z - w \right) \left( z + w \right)^\ast &=&
m_3 \left[ s^2 m_1 e^{2 i \left( \beta_1 - \beta_3 \right)}
+ c^2 m_2 e^{2 i \left( \beta_2 - \beta_3 \right)} \right].
\label{phase2}
\ea
In analogy to equation~(\ref{z-w}) we also have,
for $M_R$ instead of $\mathcal{M}_\nu$,
\be
\left| p - q \right| = M_3.
\label{p-q}
\ee

\paragraph{The parameters $a$ and $b$}
We now express $|a|$ and $|b|$
as functions of the neutrino masses. 
Using equations~(\ref{zmw}),
(\ref{z-w}) and~(\ref{p-q}) we obtain  
\be
\left| b \right|^2 = m_3 M_3.
\label{bmodulus}
\ee
Comparing $\det{\mathcal{M}_\nu}$ with $\det{M_R}$
we readily find that 
$\left| a \right|^2 \left| b \right|^4 = m_1 m_2 m_3 M_1 M_2 M_3$.
Therefore,
with equation~(\ref{bmodulus}) we conclude that
\be
\left| a \right|^2 = \frac{m_1 m_2 M_1 M_2}{m_3 M_3}.
\label{amodulus}
\ee
Moreover,
equations~(\ref{det}), (\ref{z-w}) and~(\ref{p-q}) lead to
\ba
\left| d \right| &=& M_1 M_2, \label{jgort} \label{deq} \\
\left| f \right| &=& m_1 m_2. \label{iervd}
\ea

\paragraph{The imaginary part of $\left( R_{12} \right)^2$}

From equation~(\ref{y2xzpw}) one derives
\be
2\, {\rm Im} \left[ y^2 x^\ast \left( z + w \right)^\ast \right]
= c^2 s^2 m_1 m_2 \left( m_2^2 - m_1^2 \right) \sin{\Delta}.
\label{jghuf}
\ee
Using the analogy between the matrices $\mathcal{M}_\nu$ and $M_R$,
and between $U$ and $V$,
one sees that
\be
2\, {\rm Im} \left[ n^2 m^\ast \left( p + q \right)^\ast \right]
= {c^\prime}^2 {s^\prime}^2 M_1 M_2 \left( M_2^2 - M_1^2 \right)
\sin{\left( 2 \gamma_1 - 2 \gamma_2 \right)}.
\label{nfhgq}
\ee
The matrix $R$ is given in equation~(\ref{mgowx}).
Using equation~(\ref{nfhgq}) one sees that
\be
{\rm Im} \left[ \left( R_{12} \right)^2 \right] =
\left( \left| b \right|^2 - \left| a \right|^2 \right)^2
\,
\frac{2\, {\rm Im} \left[ n^2 m^\ast \left( p + q \right)^\ast \right]}
{M_1 M_2 \left( M_2^2 - M_1^2 \right)}.
\label{uivnd}
\ee
Now,
from equations~(\ref{x})--(\ref{zpw}) one finds the relation 
\be
y^2 x^\ast \left( z + w \right)^\ast = \left| a \right|^4 \left| b \right|^4
\frac{n^2 m^\ast \left( p + q \right)^\ast}{\left| d \right|^4}.
\ee
Therefore,
equations~(\ref{jghuf}) and (\ref{uivnd}) give
\ba
{\rm Im} \left[ \left( R_{12} \right)^2 \right] &=&
\left( \left| b \right|^2 - \left| a \right|^2 \right)^2
\frac{m_1 m_2 \left( m_2^2 - m_1^2 \right)}
{M_1 M_2 \left( M_2^2 - M_1^2 \right)}\,
\frac{\left| d \right|^4}{\left| a \right|^4 \left| b \right|^4}\,
c^2 s^2 \sin{\Delta}
\label{uqwxc} \\ &=&
\left( \left| b \right|^2 - \left| a \right|^2 \right)^2
\frac{M_1 M_2 \left( m_2^2 - m_1^2 \right)}
{m_1 m_2 \left( M_2^2 - M_1^2 \right)}\,
c^2 s^2 \sin{\Delta},
\label{imimim}
\ea
where we have used equations~(\ref{bmodulus})--(\ref{deq})
in order to go from equation~(\ref{uqwxc}) to equation~(\ref{imimim}).

\paragraph{The quadratic equation
giving $\left| a \right|$ and $\left| b \right|$}
Equations~(\ref{bmodulus}) and (\ref{amodulus})
contain $m_3$ and $M_3$.
We shall now derive expressions for $|a|$ and $|b|$
which are functions solely of the observables
in equation~(\ref{physpar}).
With equations~(\ref{xabs})--(\ref{z+w}) it is easy to check that 
\be
m_1^2 + m_2^2 =
\left| x \right|^2 + 4 \left| y \right|^2 + \left| z + w \right|^2.
\ee
For $M_R$ instead of $\mathcal{M}_\nu$,
the analogous relation is
\be
M_1^2 + M_2^2 =
\left| m \right|^2 + 4 \left| n \right|^2 + \left| p + q \right|^2.
\ee
Using equations~(\ref{m})--(\ref{ppq}) and (\ref{iervd}),
we find 
\be
m_1^2 m_2^2 \left( M_1^2 + M_2^2 \right) =
\left| a \right|^4 \left| z + w \right|^2
+ 4 \left| a \right|^2 \left| b \right|^2 \left| y \right|^2
+ \left| b \right|^4 \left| x \right|^2.
\label{qtywx}
\ee
Multiplying this equation by $\left| x \right|^2 \left| b \right|^4$
and using $\left| a \right|^2 \left| b \right|^2 = m_1 m_2 M_1 M_2$
we finally obtain equations~(\ref{Beq})--(\ref{afhro}).
The solutions to equation~(\ref{afhro}) are
\be
m_1^2 m_2^2 \left| p + q \right|^2
= \left| x \right|^2 \left| b \right|^4
= \frac{- B \pm \sqrt{B^2 - 4 C}}{2},
\label{mguwa}
\ee
where we have used equations~(\ref{ppq}) and~(\ref{iervd}).
Since
$C = \left( \left| b \right|^4 \left| x \right|^2 \right)
\left( \left| a \right|^4 \left| z + w \right|^2 \right)$,
cf.\ equation~(\ref{Ceq}),
we also see that,
together with equation~(\ref{mguwa}),
\be
m_1^2 m_2^2 \left| m \right|^2
= \left| z + w \right|^2 \left| a \right|^4
= \frac{- B \mp \sqrt{B^2 - 4 C}}{2}.
\label{nrtso}
\ee
In order to determine $\theta^\prime$ one notes,
from equations~(\ref{xabs}) and (\ref{z+w}),
that
\be
\left| z + w \right|^2 - \left| x \right|^2
= \left( c^2 - s^2 \right) \left( m_2^2 - m_1^2 \right).
\ee
The analogous relation for $M_R$ is
\be
{c^\prime}^2 - {s^\prime}^2
= \frac{\left| p + q \right|^2 - \left| m \right|^2}{M_2^2 - M_1^2}
= \frac{\pm \sqrt{B^2 - 4 C}}{m_1^2 m_2^2 \left( M_2^2 - M_1^2 \right)}.
\label{iuwem}
\ee

\end{appendix}


\newpage

\begin{thebibliography}{99}

\bibitem{WMAP}
Spergel D N et al 2003
First year Wilkinson Microwave Anisotropy Probe (\textit{WMAP})
observations: determination of cosmological parameters 
\textit{Astrophys. J. Suppl.} {\bf 148} 175
(\textit{Preprint} astro-ph/0302209)

\bibitem{steigman}
Steigman G 2003
Forensic cosmology: probing baryons and neutrinos with BBN and the CBR
\textit{Preprint} hep-ph/0309347

\bibitem{KamLAND}
Eguchi K et al (KamLAND Collaboration) 2003
First results from KamLAND:
evidence for reactor anti-neutrino disappearance
\textit{Phys. Rev. Lett.} {\bf 90} 021802
(\textit{Preprint} hep-ex/0212021)

\bibitem{SNO}
Ahmed S N et al (SNO Collaboration) 2003
Measurement of the total active $^8$B solar neutrino flux at the
Sudbury Neutrino Observatory with enhanced neutral current sensitivity
\textit{Preprint} nucl-ex/0309004

\bibitem{SK}
Hayato Y 2003 
Talk given at the \textit{International Europhysics Conference
on High Energy Physics (HEP 2003)}.
Transparencies at http://eps2003.physik.rwth-aachen.de.

\bibitem{CHOOZ}
Apollonio M et al 1999
Limits on neutrino oscillations from the CHOOZ experiment 
\textit{Phys. Lett.} B \textbf{466} 415
(\textit{Preprint} hep-ex/9907037)

\bibitem{maltoni}
Maltoni M, Schwetz T, T\'ortola M A and Valle J W F 2003
Status of three-neutrino oscillations after the SNO--salt data
\textit{Phys. Rev.} D \textbf{68} 113010
(\textit{Preprint} hep-ph/0309130)

\bibitem{reviews}
Gonzalez-Garcia M C and Nir Y 2003
Neutrino masses and mixing: evidence and implications
\textit{Rev. Mod. Phys.} \textbf{75} 345
(\textit{Preprint} hep-ph/0202058) \\
Grimus W 2003
Neutrino physics -- theory 
\textit{Preprint} hep-ph/0307149 \\
Barger V, Marfatia D and Whisnant K 2003
Progress in the physics of massive neutrinos
\textit{Int. J. Mod. Phys.} E \textbf{12} 569
(\textit{Preprint} hep-ph/0308123) \\
Smirnov A Yu 2004
Neutrino physics: open questions
\textit{Int. J. Mod. Phys.} A \textbf{19} 1180
(\textit{Preprint} hep-ph/0311259)

\bibitem{yanagida}
Fukugita M and Yanagida T 1986
Baryogenesis without grand unification
\textit{Phys. Lett.} B {\bf 174} 45

\bibitem{lepto-reviews}
For reviews of leptogenesis see for instance 
Pilaftsis A 1999
Heavy Majorana neutrinos and baryogenesis 
\textit{Int. J. Mod. Phys.} A \textbf{14} 1811 
(\textit{Preprint} hep-ph/9812256) \\
Buchm\"uller W and Pl\"umacher M 2000
Neutrino masses and the baryon asymmetry
\textit{Int. J. Mod. Phys.} A \textbf{15} 5047 
(\textit{Preprint} hep-ph/0007176) \\
Paschos E A 2004
Leptogenesis
\textit{Pramana} \textbf{62} 359
(\textit{Preprint} hep-ph/0308261)

\bibitem{seesaw}
Yanagida T 1979
Horizontal gauge symmetry and masses of neutrinos
\textit{Proc. of the Workshop
on Unified Theories and Baryon Number in the Universe
(Tsukuba, Japan, 1979)}
ed O Sawada and A Sugamoto
(KEK report no 79-18, Tsukuba, 1979) \\ 
Glashow S L 1981
The future of elementary particle physics
\textit{Quarks and Leptons, Proc. of the Advanced Study Institute
(Carg\`ese, Corsica, 1979)}
ed J-L Basdevant et al
(New York: Plenum) \\
Gell-Mann M, Ramond P and Slansky R 1979 
Complex spinors and unified theories
\textit{Supergravity, Proc. of the Workshop (Stony Brook, NY, 1979)}
ed P van Nieuwenhuizen and D Z Freedman
(Amsterdam: North Holland) \\
Mohapatra R N and Senjanovi\'c G 1980
Neutrino mass and spontaneous parity violation
\textit{Phys. Rev. Lett.} {\bf 44} 912

\bibitem{schechter}
Schechter J and Valle J W F 1980
Neutrino masses in SU(2)$\, \otimes \,$U(1) theories 
\textit{Phys. Rev.} D \textbf{22} 2227 \\
See also Bilenky S M, Ho\v{s}ek J and Petcov S T 1980
On oscillations of neutrinos with Dirac and Majorana masses 
\textit{Phys. Lett.} \textbf{94B} 495 \\
Kobzarev I Yu, Martemyanov B V, Okun L B and 
Shchepkin M G 1980
The phenomenology of neutrino oscillations 
\textit{Yad. Phys.} \textbf{32} 1590 
[\textit{Sov. J. Nucl. Phys.} \textbf{32} 823]

\bibitem{fullseesaw}
Grimus W and Lavoura L 2000
The seesaw mechanism at arbitrary order:
disentangling the small scale from the large scale
\textit{J. High Energy Phys.} {\bf 0011} 042
(\textit{Preprint} hep-ph/0008179)

\bibitem{casas}
Casas J A and Ibarra A 2002
Oscillating neutrinos and $\mu \to e \gamma$ 
\textit{Nucl. Phys.} B \textbf{618} 171
(\textit{Preprint} hep-ph/0103065)

\bibitem{branco1}
Branco G C, Morozumi T, Nobre B M and Rebelo M N 2001
A bridge between CP violation at low energies and leptogenesis 
\textit{Nucl. Phys.} B \textbf{617} 475 
(\textit{Preprint} hep-ph/0107164) \\
Rebelo M N 2003
Leptogenesis without $CP$ violation at low energies 
\textit{Phys. Rev.} D \textbf{67} 013008 
(\textit{Preprint} hep-ph/0207236)

\bibitem{branco2}
Branco G C et al 2003
Minimal scenarios for leptogenesis and $CP$ violation
\textit{Phys. Rev.} D {\bf 67} 073025
(\textit{Preprint} hep-ph/0211001) \\
Siyeon K 2003
Leptogenesis and bi-unitary parametrization of the neutrino Yukawa matrix
\textit{Eur. Phys. J.} C \textbf{30} 55
(\textit{Preprint} hep-ph/0303077)

\bibitem{pascoli}
Pascoli S, Petcov S T and Rodejohann W 2003
On the connection of leptogenesis with low energy $CP$ violation and
lepton flavor violating charged lepton decays
\textit{Phys. Rev.} D \textbf{68} 093007
(\textit{Preprint} hep-ph/0302054)

\bibitem{endoh}
Endoh T, Kaneko S, Kang S K, Morozumi T and Tanimoto M 2002
$CP$ violation in neutrino oscillations and leptogenesis
\textit{Phys. Rev. Lett.} \textbf{89} 231601
(\textit{Preprint} hep-ph/0209020)

\bibitem{raidal}
Raidal M and Strumia A 2003
Predictions of the most minimal seesaw model 
\textit{Phys. Lett.} B \textbf{553} 72
(\textit{Preprint} hep-ph/0210021)

\bibitem{barger}
Barger V, Dicus D A, He H-J and Li T 2004
Structure of cosmological CP violation via neutrino seesaw
\textit{Phys. Lett.} B \textbf{583} 173
(\textit{Preprint} hep-ph/0310278)

\bibitem{felipe}
Gonz\'alez Felipe R, Joaquim F R and Nobre B M 2003
Radiatively induced leptogenesis in a minimal seesaw model
\textit{Preprint} hep-ph/0311029

\bibitem{falcone}
Falcone D 2003
Inverting the seesaw formula
\textit{Phys. Rev.} D \textbf{68} 033002
(\textit{Preprint} hep-ph/0305229)

\bibitem{akhmedov}
Akhmedov E Kh, Frigerio M and Smirnov A Yu 2003
Probing the seesaw mechanism with neutrino data and leptogenesis
\textit{J. High Energy Phys.} \textbf{0309} 021
(\textit{Preprint} hep-ph/0305322)

\bibitem{velasco}
Velasco-Sevilla L 2003 
Hierarchical neutrino mass matrices, CP violation and leptogenesis 
\textit{J. High Energy Phys.} \textbf{0310} 035
(\textit{Preprint} hep-ph/0307071)

\bibitem{rodejohann}
Rodejohann W 2004
Hierarchical matrices in the seesaw mechanism,
large neutrino mixing and leptogenesis
\textit{Eur. Phys. J.} C \textbf{32} 235
(\textit{Preprint} hep-ph/0311142)

\bibitem{Z2}
Grimus W and Lavoura L 2001
Softly broken lepton numbers and maximal neutrino mixing
\textit{J. High Energy Phys.} {\bf 0107} 045
(\textit{Preprint} hep-ph/0105212) \\
Grimus W and Lavoura L 2001
Softly broken lepton numbers: an approach to maximal neutrino mixing
\textit{Acta Phys. Polon.} B {\bf 32} 3719
(\textit{Preprint} hep-ph/0110041)

\bibitem{D4}
Grimus W and Lavoura L 2003
A discrete symmetry group for maximal atmospheric neutrino mixing
\textit{Phys. Lett.} B {\bf 572} 189
(\textit{Preprint} hep-ph/0305046)

\bibitem{review-models}
Grimus W and Lavoura L 2003
Models of maximal atmospheric neutrino mixing
\textit{Acta Phys. Polon.} B \textbf{34} 5393
(\textit{Preprint} hep-ph/0310050)

\bibitem{LB}
Luty M A 1992
Baryogenesis via leptogenesis
\textit{Phys. Rev.} D \textbf{45} 455 \\
Flanz M, Paschos E A and Sarkar U 1995
Baryogenesis from a lepton asymmetric universe
\textit{Phys. Lett.} B \textbf{345} 248 
(\textit{Preprint} hep-ph/9411366) \\
Pl\"umacher M 1997
Baryogenesis and lepton number violation 
\textit{Z. Phys.} C \textbf{74} 549 
(\textit{Preprint} hep-ph/9604229) \\
Covi L, Roulet E and Vissani F 1996
CP violating decays in leptogenesis scenarios 
\textit{Phys. Lett.} B \textbf{384} 169 
(\textit{Preprint} hep-ph/9605319) \\
Buchm\"uller W and Pl\"umacher M 1998 
CP asymmetry in Majorana neutrino decays 
\textit{Phys. Lett.} B \textbf{431} 354 
(\textit{Preprint} hep-ph/9710460)

\bibitem{takanishi}
Nielsen H B and Takanishi Y 2001
Baryogenesis via lepton number violation in anti-GUT model
\textit{Phys. Lett.} B {\bf 507} 241
(\textit{Preprint} hep-ph/0101307) 

\bibitem{pilaftsis}
Pilaftsis A and Underwood T E J 2003
Resonant leptogenesis
\textit{Preprint} hep-ph/0309342

\bibitem{roos}
Roos M 1994 
\textit{Introduction to Cosmology} 
(John Wiley \& Sons)

\bibitem{harvey}
Harvey J A and Turner M S 1990
Cosmological baryon and lepton number in the presence of electroweak
fermion number violation  
\textit{Phys. Rev.} D \textbf{42} 3344

\bibitem{buchmueller}
Buchm\"uller W, Di Bari P and Pl\"umacher M 2002
Cosmic microwave background,
matter-antimatter asymmetry and neutrino masses 
\textit{Nucl. Phys.} B \textbf{643} 367 
(\textit{Preprint} hep-ph/0205349)

\bibitem{kolb}
Kolb E W and Turner M S 1990
\textit{The Early Universe} (Addison--Wesley)

\bibitem{notari}
Giudice G F, Notari A, Raidal M, Riotto A and Strumia A 2004 
Towards a complete theory of thermal leptogenesis in the SM and MSSM 
\textit{Nucl. Phys.} B \textbf{686} 89
(\textit{Preprint} hep-ph/0310123)

\bibitem{klapdor}
Klapdor-Kleingrothaus H V, Dietz A, Harney H L
and Krivosheina I V 2001
Evidence for neutrinoless double beta decay 
\textit{Mod. Phys. Lett.} A \textbf{16} 2409
(\textit{Preprint} hep-ph/0201231)

\end{thebibliography}
\end{document}